\documentclass[10pt]{template-SCTC}

\ifCLASSINFOpdf
  \usepackage[pdftex]{graphicx}
\else
  \usepackage[dvips]{graphicx}
\fi

\usepackage{multirow}
\usepackage{multicol}
\usepackage{subfigure}	
\usepackage{flushend}
\usepackage[T1]{fontenc}
\usepackage[utf8]{inputenc}
\usepackage{enumitem}
\setlist[itemize,1]{leftmargin=0.25in}

\usepackage[draft]{hyperref}

\usepackage{balance}

\begin{document}
\newcommand*{\nextpar}{\vspace{6.0pt}} 
\pagenumbering{gobble}  

\twocolumn[
\begin{@twocolumnfalse}

\vspace{10pt} \noindent

\begin{center}
{\fontsize{20}{24}\selectfont
HoloMed: A Low-Cost Gesture-Based Holographic}

\vspace{0.30cm}
{\fontsize{20}{24}\selectfont
System to Learn Normal Delivery Process}
\vspace{0.2cm}

Juan Perozo$^{1}$, Mimia Lo Leung$^{1}$, Esmitt Ramírez$^{2}$\\
jperozo89@gmail.com, mimia.loleung@gmail.com, esmitt.ramirez@ciens.ucv.ve \\
\vspace{0.4cm}

$^{1 \:}$Computer Engineering School, Andrés Bello Catholic University. Caracas, Venezuela\\
$^{2 \:}$Computer Science School, Central University of Venezuela. Caracas, Venezuela
\end{center}

\begin{center}
\noindent\rule{510pt}{0.4pt}
\end{center}
\vspace{-0.1cm}

\begin{center}
\begin{minipage}{0.95\textwidth}
\textbf{Abstract:} During medicine studies, visualization of certain elements is common and indispensable in order to get more information about the way they work. Currently, we resort to the use of photographs -which are insufficient due to being static- or tests in patients, which can be invasive or even risky. Therefore, a low-cost approach is proposed by using a 3D visualization. This paper presents a holographic system built with low-cost materials for teaching obstetrics, where student interaction is performed by using voice and gestures. Our solution, which we called HoloMed, is focused on the projection of a  euthocic normal delivery under a web-based infrastructure which also employs a Kinect. HoloMed is divided in three (3) essential modules: a gesture analyzer, a data server, and a holographic projection architecture, which can be executed in several interconnected computers using different network protocols. Tests used for determining the user's position, illumination factors, and response times, demonstrate HoloMed's effectiveness as a low-cost system for teaching, using a natural user interface and 3D images.
\nextpar


\end{minipage}
\end{center}
\vspace{-0.1cm}
\begin{center}
\noindent\rule{510pt}{0.4pt}
\end{center}

\end{@twocolumnfalse} ]

\section{Introduction}

Childbirth represents the end of a formation process which begins when a spermatozoid fertilizes an egg. This process takes an average time between seven and nine months. This is a normal behavior of our species, but it is not free of risks.3. According to a report of the World Health Organization from 2015 \cite{data}, over 830 women die every day due to causes related to pregnancy and childbirth that can be prevented. So, it becomes crucial for obstetricians to be in their full capabilities in order to identify if there could be risk for the child and the mother during labor.
\nextpar

Medicine's major involves a huge amount of information condensed in thousands of images and books. Nowadays, students must apply a mental spatial reconstruction for some angles in some particular set of images. Practicing with plastic models or real patients are mandatory in order to get a full understanding of the topic being learned. Childbirth or child delivery is one of the most difficult processes when it comes to capturing images from it. The physicians focused on the study of pregnancy, childbirth and the postpartum period are called obstetricians.
\nextpar

In Obstetrics, the development of the fetus is the part of the studies where the size and the growth of the fetus are considered.. The culmination of the period of pregnancy is known as childbirth, where the fetus is expelled from the uterus of a woman (i.e. natural delivery). When the fetus is in a longitudinal position and the head enters the pelvis first, is called a cephalic presentation. This presentation, also known as head-first presentation, is the most common in normal delivery, which occupies the 96.5\% of cases. Other kinds of presentations are called podalic and abnormal \cite{reilly}.
\nextpar

\begin{figure}[!htpb]
  \includegraphics[width=\columnwidth]{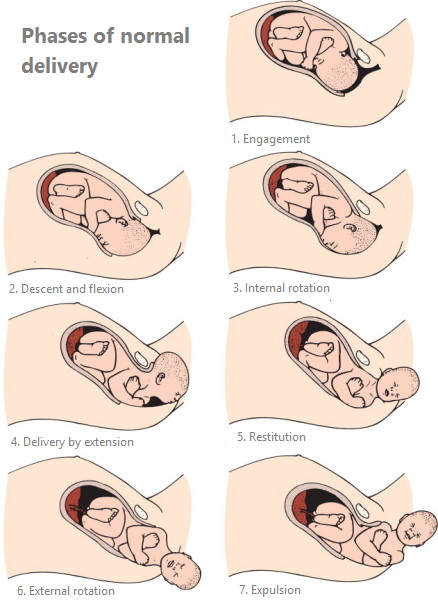}
  \caption{Graphical representation of the phases of normal delivery.}
  \label{fig:phases}
\end{figure}

A delivery is considered as normal when there is no complication during the childbirth. The head and shoulders of the infant must follow a specific sequence of maneuvers in order to pass through the ring of the pelvis. The process is considered to have seven phases: engagement, descent, flexion, internal rotation, delivery by extension, restitution, external rotation and expulsion. As mentioned before, phases of natural delivery are totally visual according to the spatial location of the newborn and the mother (see Figure \ref{fig:phases}).
\nextpar

The medical sector is usually at the forefront of technological innovations. Nowadays, there are several solutions in order to display more realistic medical images using new technologies such as 3D images. Often, these solutions are part of daily practice in health-care facilities. However, during the Medicine major, those solutions are not present due to their costs, setup, and availability in the higher education centers. A novel solution in medical images is using holograms, which allow a 3D graphical representation from several points of view of a region of interest of the body.. 
\nextpar

Using holograms to pop-up images from books will be an extraordinary improvement to learn Medicine. In this research, we present a solution which intends to serve as an educational platform to teach the head-first presentation childbirth by normal delivery showing in 3D images. Our solution presents all the phases of the fetus during the birth process, letting the students interact using natural interfaces based on gestures.
\nextpar

HoloMed, derived from \textit{Hologram for Medicine}, is the name of our solution. HoloMed is a complete solution where teachers can enter into a website to write contents, questions and answers about the process and set a group of images to display the different phases as 3D floating objects. Students can answer the questions using gestures captured by a Kinect to gain points. Also, it gives and takes voice commands to increase the interactivity. The architecture of HoloMed is based on a real-time network communication using a web browser to be executed independently of the operating system.
\nextpar

\section{Related Work}	\label{sec:relatedwork}

In Science, the visualization of the studies elements is mandatory and with the use of laboratories or open field studies to verify some hypotheses this can be perfectly possible \cite{vis}. Taking pictures and recording videos are common practices in order to store permanent data for future studies. However, they lack an important aspect: 3D spatial reconstruction of the images captured. Of course, there are 3D camcorders in the market, but those are expensive and almost new to establish them as a fixed equipment in Science classes.
\nextpar

In 1947 Dennis Gabor invented the concept of holographic method, getting the Nobel Prize in Physics in 1971 \cite{nobel} for this achievement. A hologram is a photographic emulsion where the information about the scene is recorded in a special way \cite{holo1}. When the hologram is properly illuminated, the viewer sees a realistic 3D representation of the scene. Its usage is widespread for cultural, educational and commercial purposes \cite{holo2}.
\nextpar

\begin{figure}[!htpb]
  \includegraphics[width=\columnwidth]{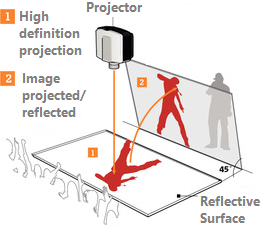}
  \caption{Scheme of the holography presented in the Coachella Music Festival in 2012 \cite{coachella}.}
  \label{fig:tupac}
\end{figure}

In the entertainment world, a remarkable example of holography was the stage performance of singer Tupac Shakur, died in 1996, with Snoop Dog in the Coachella Music Festival \cite{coachella}. This was achieved by using a Pepper's ghost illusion which is a holographic technique. Figure \ref{fig:tupac} shows a scheme of the technique used in Coachella, where the red figure represents the projected hologram.
\nextpar

\begin{figure*}[!htpb]
  \includegraphics[width=\textwidth]{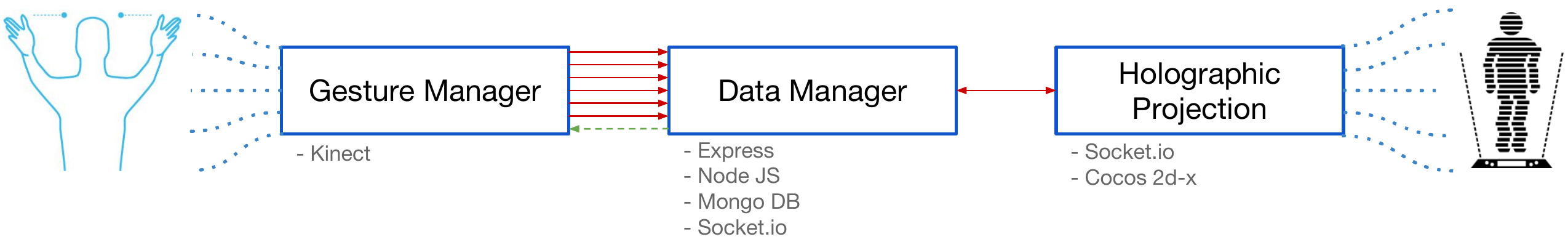}
  \caption{General overview of our proposal showing the three modules involved, plus the input (left) and output (right).}
  \label{fig:overview}
\end{figure*}

According to Mehta \cite{recent}, recent improvements in hologram are based on recording techniques and applied as tools for the success of holographic imaging through tissues, ophthalmology, dentistry, urology, pathology and orthopedics. This shows a strong promise for holography to emerge as a powerful tool for medical applications.
\nextpar

In 2013, Hackett \cite{hackett} proved that medical holograms provide a statistically significant performance improvement over traditional textbook materials in understanding anatomy, especially in regards to spatial relationships. Also, the study deepens into the cognitive load measures were lower using the holograms as well, lending credence to the notion that it is easier and more intuitive to process complex medical information from holograms.
\nextpar

There are a few companies dedicated to creating more realism to the given medical data. Regardless if output is a 2D monitor, the idea is to create some effect similar to 3D movies or consoles such as Nintendo 3DS. Noticeable examples are the companies EchoPixel \cite{echopixel}, Zebra Imaging \cite{zebra} and RealView Imaging \cite{realview}. For instance, the software True 3D Viewer developed by EchoPixel \cite{echopixel} is a real-time interactive virtual reality system integrated for DICOM data, using a hand-directed stylus as if they were real physical objects.
\nextpar 

Furthermore, there are other companies focused on creating floating human body parts. A clear example is NanoLive \cite{nano}, which presents the product 3D Cell Explorer. This works as a microscope that uses holographic to create detailed stereoscopic visualizations of microbial and cellular life. It shows a combination of holographic and rotational scanning to detect changes of light as it propagates through a cell.
\nextpar

There are a full set of techniques to perform holography. Also, its application depends on the cover area. Thus, a basic introduction of techniques is explained in \cite{svodo} where authors presented an approach for in-house devices. This aspect is a keystone in the development of holographic hardware and software: their cost. Generally, the equipment is too expensive to be acquired by institutions. Nevertheless, there are options to construct low-cost solutions following the basic techniques. This way, given the impact and the momentum of holography in the modern world as a visual communication tool, our research is based on this topic.

\section{Overview}	\label{sec:ourapproach}

As mentioned before, HoloMed offers a solution to display a hologram for a normal delivery on its different stages. HoloMed arises as a teaching support for the physicians training with a modern 3D interface, getting the based-gestures input from the user using a Kinect. A final user (i.e. students) has to answer some questions while watching the hologram in the visualization pyramid constructed by us. This question was previously written by the administrator user (i.e. teacher) and answered by the students by using gestures captured by the Kinect. The hologram is shown over a holographic visualization pyramid, which shows the fetus on each stage, according to the progress of correct answers.
\nextpar

Figure \ref{fig:overview} shows a general overview of HoloMed. It is divided into three modules: Gesture Manager, Data Manager, and Holographic Projection. The Gesture Manager is able to capture and process gestures performed by final users. This information is sent to the Data Manager module, which is responsible to take actions using the input information, in the whole system. This module controls input/output in HoloMed, and it stores data in a document-oriented database. Finally, the response to user inputs is achieved in the Holographic Projection module. It contains the holographic visualization pyramid architecture.
\nextpar

It is important to note that each module might be connected in a different machine (e.g. PC, tablet or mobile device). The Device Manager receives the gesture information from Gesture Manager, and sends instructions about the image to draw to the Holographic Projection. All modules must be synchronized before setting or getting any data. The initial setup is a keystone to get right movements.
\nextpar

Data Manager uses the Socket.IO library \cite{socket} to get a real-time communication between user input and projected hologram. For instance, this module requires supports of the Mongo DB database \cite{mongo}, Node.js \cite{node} and Express \cite{express} libraries to run HoloMed over a web browser. Similarly, output architecture needs support of Socket.io and Cocos2D-x \cite{cocos} as visualization library. Indeed, a tablet with any video output is enough to connect the holographic visualization pyramid.
\nextpar

The output presented is based on a fetus on its different phases in the normal delivery process. This is previously constructed by using a set of Spritesheets. A Spritesheet is a number of ordered images of fetus developed to be displayed by the holographic visualization pyramid. The section \ref{sec:spritesheet} explains this aspect in detail. Following, the web architecture placed on the Data Manager module is shown in section \ref{sec:webarchitecture}. The remaining modules, Gesture Manager and Holographic Projection are presented in sections \ref{sec:gesture} and \ref{sec:holo} respectively.

\section{SpriteSheet}	\label{sec:spritesheet}

The 3D model built corresponds to a formed fetus with 37 to 42 weeks of prenatal development. It was developed by using Blender \cite{blender} as a capsule structure which contains the skeleton with proper controllers to reach movements of character. The skeleton defined uses rigging, allowing a deformation degree due to the weight of each composed vertex. Each weight impacts over the full movement of the fetus.
\nextpar
\begin{figure}[!htpb]
  \includegraphics[width=0.9\columnwidth]{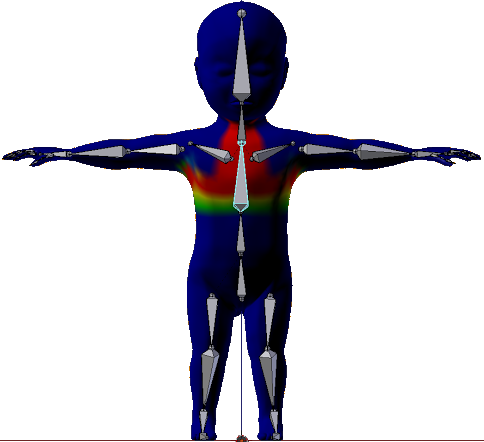}
  \caption{3D model of fetus used showing rigging and weight features.}
  \label{fig:vertice}
\end{figure}

Figure \ref{fig:vertice} shows a representation of the constructed model. Note the rigging structure over the skeleton (gray color skeleton). The colors represent the weight of each part of the fetus according to its impact over the deformation of the object. The red color area located in the chest of the model indicates the higher weight or zones with major influence, while the lower weight or minor influence are represented as a degradation to blue color. These colors are based on the Weight Painting Mode provided by Blender.
\nextpar

\begin{figure}[!htpb]
  \includegraphics[width=\columnwidth]{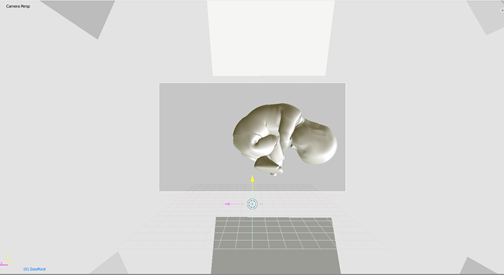}
  \caption{Simulation in Blender of the camera in frontal view of the fetus in Phase 1 (engagement).}
  \label{fig:fetus2}
\end{figure}

Once the model is created, it must be tested properly. A simulation of the holographic system is performed by placing a camera aligned with the X-axis to spin around the fetus. The reached effect is a rotary clockwise movement simulating the illusion that the fetus is rotating over its center (over its Y-axis). A representation in Blender of this is shown in Figure \ref{fig:fetus2}.
\nextpar

\begin{figure}[!htpb]
  \includegraphics[width=\columnwidth]{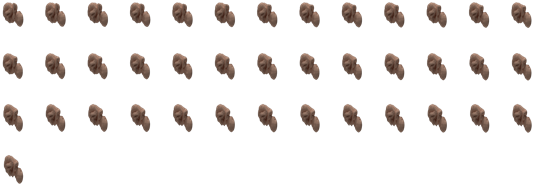}
  \caption{An example of a spritesheet formed by 40 images.}
  \label{fig:spritesheet}
\end{figure}

When all captured images are set together in one single image, for efficiency issues, they are called spritesheet (see Figure \ref{fig:spritesheet}). Camera is placed in front, lateral and posterior views to generate a spritesheet. Each spritesheet represents each phase, with 40 images each. Then, a total of $40$ images $\times$ $7$ phases $= 280$ images were generated, plus 1 image for the final phase, for a total of 281 images divided in 8 spritesheets.
\nextpar

The importance of the sprisheet is to standardize the model of fetus and being compatible with low-end devices. It is important to indicate the model was developed in several iterations in order to fit according specifications written in the anatomy book of Aller and Pagés \cite{obstetricia}. Also, we followed the feedback of supported physicians for this research. Moreover, a mechanism to map the controllers with bones was developed due to the anatomical movements which are closer to real ones.
\nextpar

\section{Web Architecture}	\label{sec:webarchitecture}

The web architecture is based on a central manager, which is able to group functionalities for students, contents, answers, questions, holograms, and NUIs (natural user interfaces). Specifically, the hologram group controls the display options to the Holographic Projection module: size, intensity, angle of projection, and others. The NUI allows the configuration of gestures in order to associate a gesture with an answer. For instance, move the right arm from left to right indicates ``Yes'', and from left to right indicates ``No''.
\nextpar

All groups in the web architecture contain the classic CRUD operations based on the Express.js web framework. This framework is a reduced and flexible web application framework for Node.js, which is an event-driven, non-blocking I/O model of web server in Javascript. Node.js is efficient because it is developed under the V8 Javascript engine of Chrome, which is fast and lightweight for real-time responses. Those tools were used to create the back-end or internal support for all funcionalities. Also, the back-end manages the complete database accesses. 
\nextpar

The database is created to store the information of teachers, students, questions, answers and phases of the fetus. To obtain a real-time response of the Data Manager module to the Gesture Manager module, the database was constructed using a document-oriented approach using Mongo DB. The books stored the information corresponding to any session, where all actors mentioned before are involved (e.g. teachers, students, contents, questions, and so on). The input data is provided by teachers, and consulted by students.
\nextpar

The front-end or visual interface available to administrator users (i.e. teachers) was developed thinking in its usage in any device. For this, JQuery \cite{jquery} and Bootstrap \cite{boot} were used, given that they have proper support in several web browsers.

\subsection{Communication}

Considering that Holographic Projection (visualization pyramid), Gesture Manager (Kinect) and Data Manager (database and web platform) are independent from each other, it is mandatory 2. a network design to support their real-time communication is mandatory.
\nextpar

The communication between Data Manager and Gesture Manager modules must be in real time. Our approach uses Socket.io which is a JavaScript library for real-time web applications which allows a bi-directional communication between web clients and servers. This library uses the WebSocket protocol with polling as a fallback option using the same programming interface.
\nextpar

Figure 3 shows a representation of the communication between both Gesture Manager and Data Manager modules. Note arrows from left to right (solid red color) and the arrow from right to left (dashed green color). Those represent the amount of load data passed from each module to the other. Also, the bidirectional communication of images to be rendered into the Holographic Projection is an open channel under the Web Socket protocol.
\nextpar

When a gesture is detected, movements are sent to the Data Manager as a HTTP request. The Data Manager verifies the validity of gesture and sends their evaluation to the Holographic Projection module, which will produce a final image in the visualization pyramid (i.e. hologram).
\nextpar

The Holographic Projection module always sends and receives data from Data Manager to establish a full stream based on control of rendered images. Also, any error will be notified immediately to final users.
\nextpar

\section{Gesture Manager}	\label{sec:gesture}

This module is the responsible of capturing data from the Kinect. There is an algorithm written in Python which encapsulates gestures as one single HTTP request to the Data Manager. This script is written using a library named libfreenect, developed by the OpenKinect open community \cite{openkinect}. The library allows the capturing and management of the data received from the Kinect by several operating systems. For each gesture, there are 3 attempts before output an error in the communication.
\nextpar

The students interact with HoloMed using only single movements. Those are explained before the start of a lesson. Then, each student must log into the platform using a keyboard/mouse with a classic application session. A session is created to study a phase in the natural delivery process. Answers are defined as true or false, this offers an easy interaction due to being gesture-based on this point. Also, actions such as switching between lessons, session log off, check hint and others, are performed using gestures too.
\nextpar

\begin{figure}[!htpb]
  \includegraphics[width=\columnwidth]{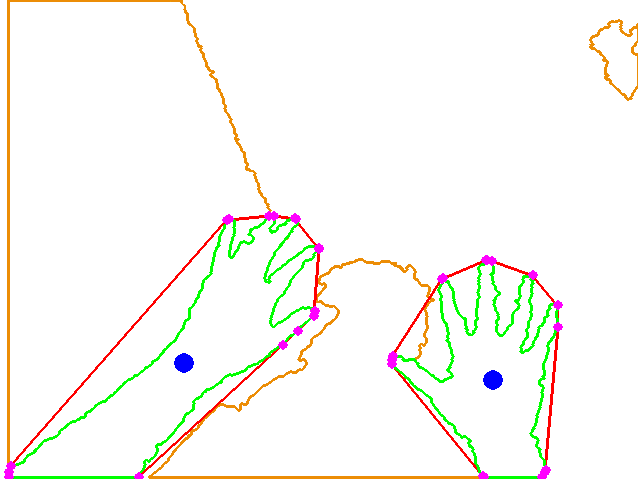}
  \caption{An instant of time $t$ of the captured image with the Kinect.}
  \label{fig:kinect}
\end{figure}

An example of a captured and processed image is shown in Figure \ref{fig:kinect}. The noise in the image is cleaned, isolating the user (i.e. applying filtering algorithms), and only the contour/border is considered. The blue points represent a reference location of the hand in time $t$. This determines the relative position of a hand in a frame. In order to avoid noise, we obtained the frame in $t-1$ and $t+1$ joined with frame in $t$ to average the relative position according to the coordinate system established in that session.

It is clear that communication between a student user with HoloMed is using gestures. However, the response of HoloMed to student user during the lessons is using spoken words. In fact, there are some commands from user to the system using command voices. The library ResponsiveVoice.js \cite{response} was added to obtain this feature in web pages. The spoken data is extracted directly from the Data Manager module, requesting some information such as description of a movement of the fetus in a particularly stage.

\section{Holographic Projection}	\label{sec:holo}

An important part of the Holographic System module is the construction of the visualization pyramid using low-cost materials. The goal is to make the visualization environment without specialized materials to display the ``floating'' fetus. For this, the Gaussian holographic optical is used taking into account the reflection of light with reflective materials.
\nextpar

\begin{figure}[!htpb]
  \includegraphics[width=\columnwidth]{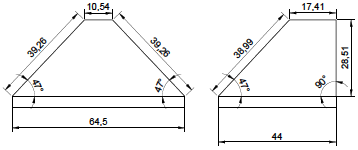}
  \caption{Blueprint of the visualization pyramid constructed.}
  \label{fig:pyramid}
\end{figure}

The first step was the design of the pyramid, creating a blueprint before its construction. Figure \ref{fig:pyramid} presents the final design, after 3 previous designs, performed using AutoCAD software. All measures shown are in centimeters.
\nextpar

The blueprint of Figure \ref{fig:pyramid} shows an angle of $47^\circ$ where the location of the projector must be at the bottom of this structure. The projector used is a conventional 21 inches computer monitor, preferable flat screen. That is a design reason for proportion of pyramid. Also, the usage of $47^\circ$ instead of the classic $45^\circ$ is to choose the best projection of images over the photo-sensitivity material used. However, images to be render have to be introduced with a small perspective correction to fix the tilt of the pyramid.
\nextpar

After the construction of the visualization pyramid, the real-time communication between this module and the Gesture Manager must be considered, in order to achieve an adequate response from the hologram to the movement performed by the final user.
\nextpar

In the rendering process, the library Cocos2D-x \cite{cocos} was used to accomplish the display of the fetus spritesheet in the pyramid. Cocos2D-x is an open-source cross platform for game development, being flexible, lightweight and effective to expected render. Thus, spritesheets to be projected are classified according to each face of the visualization environment.
\nextpar

\begin{figure}[!htpb]
\centering
  \includegraphics[width=\columnwidth]{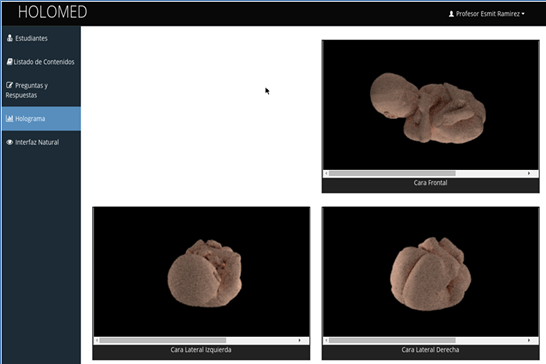}
  \caption{Web interface presented to teachers where the hologram is shown.}
  \label{fig:webgui}
\end{figure}

As mentioned before, a spritesheet contains 40 sprites where each one represents a distinct movement on a phase. This rule is applied for each cardinal point of the hologram, associating faces of the visualization pyramid with it. It is worth noting that the images are shown both in the visualization pyramid and the web interface for teachers, as presented in Figure \ref{fig:webgui}. With this, it is possible to verify the progress of students. HoloMed achieves an approximate synchronization of 25 to 30 frames per seconds in average. With this, a smooth non-perceptible transition is reached for the hologram.
\nextpar

After the full description of each stage of HoloMed, we performed a set of tests to prove our system in the learning process on the normal delivery for first-head presentation.

\section{Tests and Results}	\label{sec:evaluation}

The system was tested over different operating systems for each module, using Windows and Linux indistinctly. The higher performance used was an Intel Core Duo, with a processor of 3 GHz and a memory of 2 Gb. The monitor of the visualization pyramid was a Benq LCD monitor of 21 inches.
\nextpar

Following, we present the experiments achieved in order to demostrate its usage as a support learning tool for the natural delivery process.
\nextpar

\subsubsection{Simulation}
The simulation generated for the natural delivery is entirely focused on the fetus without considering the mother. Following, the description of each stage included in simulation:
\begin{enumerate}
    \item Floating head before fetus descending.
    \item Fetus descending until reaching trunk, forcing the flexion of the fetus's head. With this movement, the chin touches the sternal notch.
    \item The head of the fetus fit in the cervix of the mother.
    \item The fetus continues to descend in an oblique way through the pelvis where it rotates.
    \item Extension of fetus through the hole of the vulva.
    \item A rotation performed to enclose shoulders, doing the same movements of the head to be prepared for the next change.
    \item Fetus is almost expelled, output shoulders, first the right shoulder then the left one, finishing with the head.
    \item To finally expell the fetus, the head goes first, then the shoulders and the rest of the body.
\end{enumerate}

The simulation allows the visualization of defined movements on each spritesheet with dynamic gesture-based interaction. Also, the auditive output concerns about the natural user interface with the final user being an important feature in HoloMed.
\nextpar

\subsubsection{Distance Kinect-User}

To achieve a maximum performance in different architectures, we tested several configurations. For example, placing the Kinect below/above the pyramid with distances from 40 to 150 centimeters. After response tests, we found a proper distance between the user and the Kinect: between 70 and 80 centimeters. All lower measures could cause high appearance of false-positive answers to the system. Oppositely, bigger distances could confuse the readings, based on that algorithm ignores objects from a certain range.
\nextpar
\subsubsection{Lighting conditions}

\begin{figure}[!htpb]
\centering
\subfigure[Light with an approximate between 250-350 lumens. ]{\includegraphics[width=0.95\columnwidth]{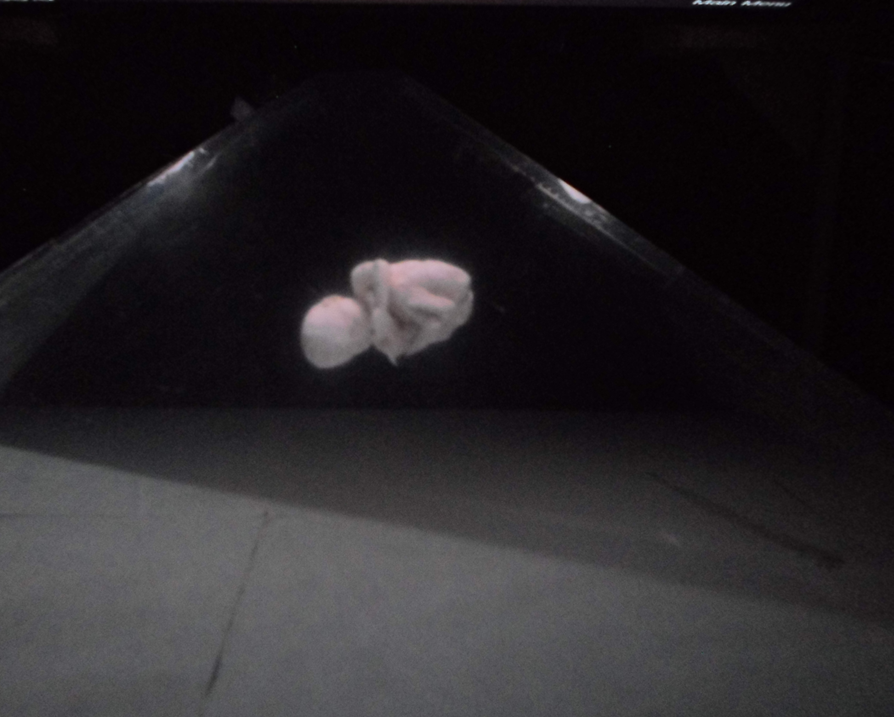}}
\subfigure[Light with an approximate between 1400-1600 lumens.]{\includegraphics[width=0.95\columnwidth]{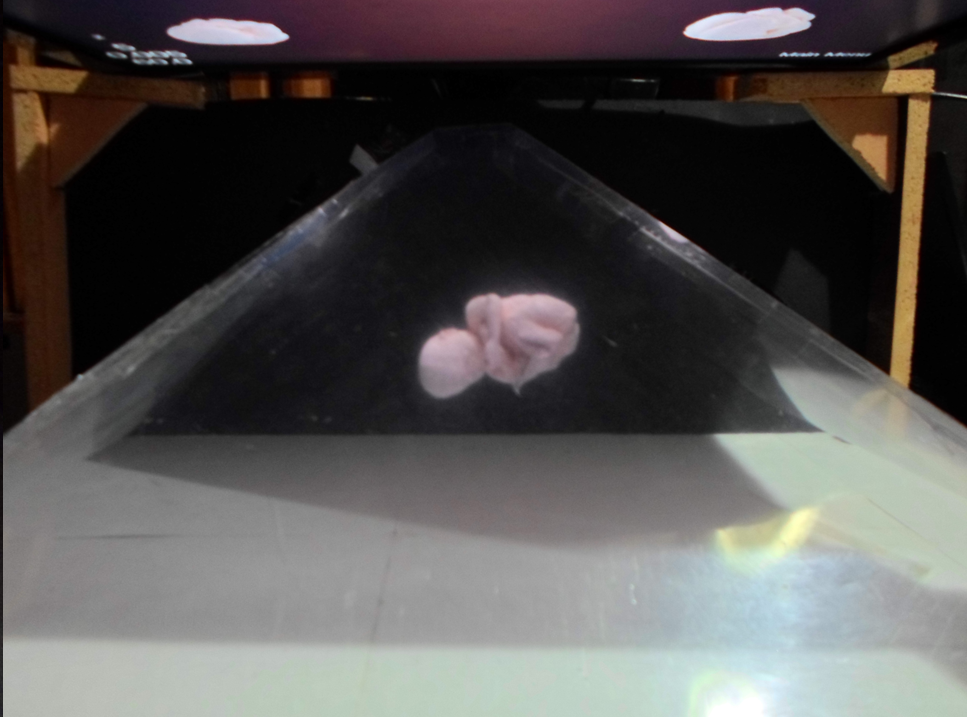}}
\caption{Ambiental condition in a room of $5$ m$^2$} \label{fig:light}
\end{figure}

A relevant test is about the lighting condition where the visualization pyramid lays. As shown on Figure \ref{fig:light}, the ambient lighting does not have a significant impact over the quality of the reflected object in the holographic architecture that was built, as long as it is not a source of total light.
\nextpar

Note that we evaluate only ambient light. It is expected that any direct light over any surface of the pyramid (e.g. a spotlight) will affect the visual perception to users.
\nextpar
\subsubsection{Response Time}

\begin{figure}[!htpb]
  \includegraphics[width=\columnwidth]{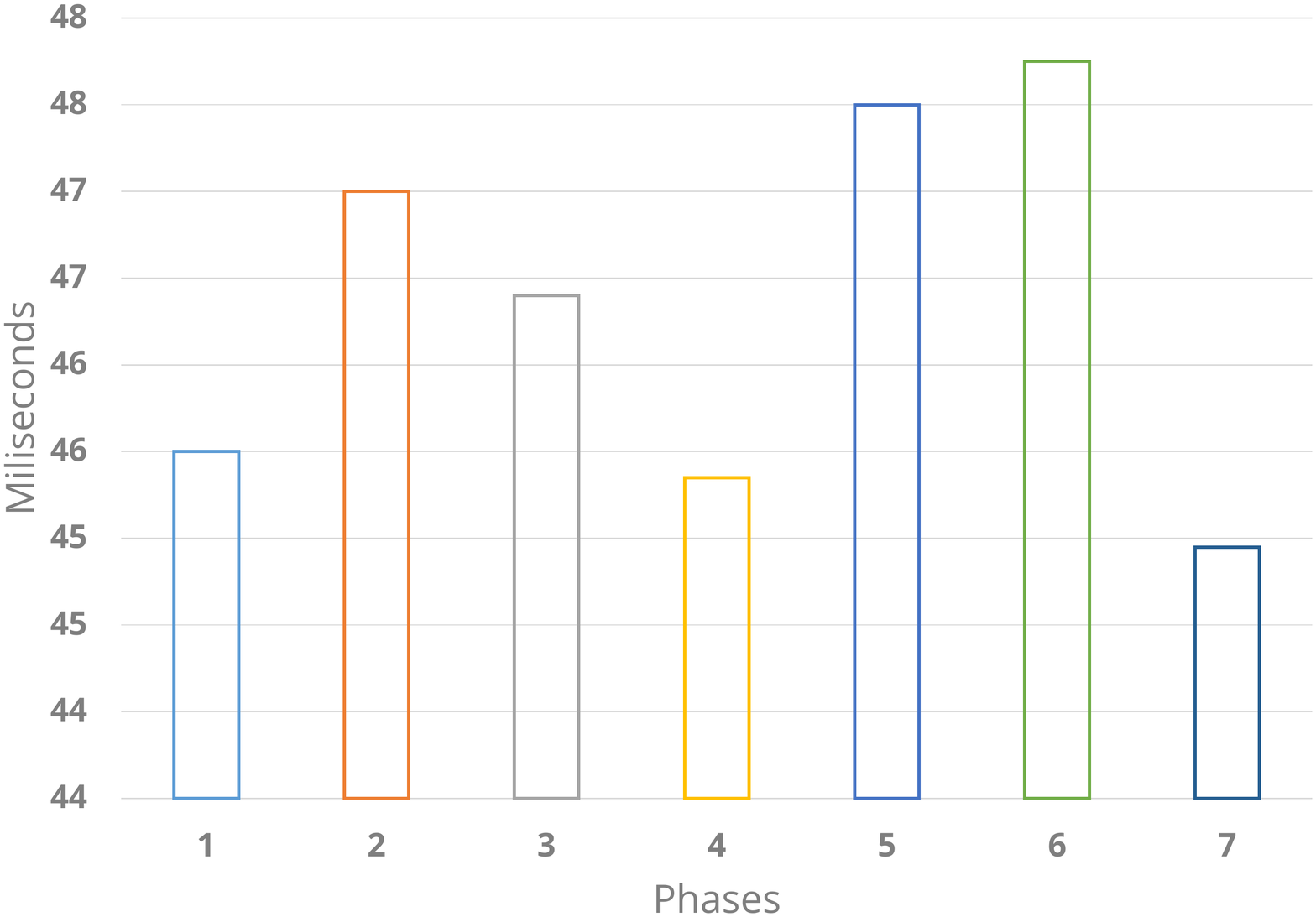}
  \caption{Graphical representation of the average time of response on each phase of normal delivery.}
  \label{fig:chart1}
\end{figure}

Figure \ref{fig:chart1} represents the average time since a signal is captured by the Gesture Manager until its representation in the Holographic Projection display. It represents, each phase of normal delivery (X-axis) with the time measure in milliseconds (Y-axis).
\nextpar

Note that all times oscillate between 41 to 58 milliseconds, with an average time of 46 milliseconds. This is considered a low value, considering a LAN network configuration of 10 Mbps with 5 machines connected. However, using an Internet connection with an ADSL connection (1 Mbps download, 0.5 Mbps uploading) we reached times up to $2.5x$ slower. It represents a considerable impact factor.
\nextpar

Additionally, we studied the average time between a gesture and the evaluation of a question. We found a range from 78 to 92 milliseconds, obtaining an average time of 80 milliseconds. The results could differ from one architecture to another, based on external elements such as the computing power of modules to network connection.
\nextpar

\subsubsection{Limitations}	\label{sec:limitation}

A classic home-made holographic system counts with a laser over a photosynthetic screen. Then, in order to use low-cost materials, we used a regular monitor over a cardboard visualization pyramid. This under controlled ambient lighting conditions. Also, each module hardware might be a low-end conventional machine. In this way, some limitations in HoloMed are given by:
\nextpar

\begin{itemize}
    \item Gesture Manager and Holographic Projector must run in a HTML5-support browser.
    \item The Holographic Projection requires a 21 inches monitor. If the size changes, the pyramid visualization must be reconstructed to adjust its size.
    \item The reflective material for the visualization pyramid is based on acetate sheets.
    \item The output device for the pyramid could be a tablet with a proper connector to the monitor (e.g. mini HDMI, display port, etc). Also, the tablet must be connected to any connected network for communication with the Data Manager module.
    \item The rotation of the hologram will always be on the Y-axis of the fetus, which means over its cardinal directions.
\end{itemize}
\nextpar

Despite of different kind of normal delivery, in this research only cephalic presentation is considered due to its frequency on normal delivery process.

\section{Conclusion and Future Work} \label{sec:conclusions}

In this investigation, the aim was to propose a visual holography-based tool to support the learning of normal delivery process. This tool is called HoloMed, based on a web architecture to allow its functionalities over any HTML5-based web browser.
\nextpar

The network infrastructure offers an important decoupling level for each stage. It allows being substituted by newer versions without having an impact on the whole system, as well as allowing distribution of computing across different devices. Also, the software created is flexible to be upgraded or replaced by others in order to improve the performance and quality of output.
\nextpar

The spritesheets, limits the fetus movements on a frame sequence, presenting a significant memory thrift. This solution has to deal with others alternatives which can work directly with the object (i.e. 3D render). However, these alternatives might cost high computing resources (space and time), since images must be built in real-time.
\nextpar

As shown in results, the ambient illumination does not have a considerable impact over the quality of the object reflected by the holographic architecture. Nevertheless, the quality is affected by the size of the visualization pyramid. This will carry the dimensions problem in blueprint, causing changes on its design to increase the size of monitor.
\nextpar

Usage of spoken words to the interactivity eliminates the need to store voice as files, since it is sent directly over the open channel using the websocket protocol. Allowing the interpretation of text in real time. The only downside of this approach is the required Internet connection.
\nextpar

This research reflects the building of an interactive hologram, using low-cost materials, and a network infrastructure which contributes to the integration between user commands. Also, in our investigation we did not find a similar system developed in Venezuela, being a novelty approach in this field of study in our country for educational, commercials or advertisement purposes.
\nextpar

This research has given many questions in need of further investigation. First, using different materials in visualization pyramid will allow a more high-quality ended product instead of a prototype. Next, we will integrate the front-end interface with a previous research \cite{ramirez12} to test other kind of medical images as input. Similarly, we desire to unify all platforms in order to use a web-based stack of development including the Kinect \cite{moreno15}.

\balance

\section*{Acknowledgment}
Authors would like to thank Md. Yuen Yee Lo and Md. Aurelysmar Griman for their valuable input and feedback for the development of this research. Also, the authors want to thank Luis Molina and Jean Cheng by their support on the pyramid's construction.



\mbox{ }
\end{document}